\documentclass[structabstract]{aa}  
\usepackage{graphicx}
\usepackage{natbib}
\usepackage{afterpage}
\bibpunct{(}{)}{;}{a}{}{,} 
\usepackage{amssymb}
\usepackage{txfonts}
\begin{document}
   \title{A representative sample of Be stars V: H$\alpha$ variability}

   \subtitle{}

   \author{R.M.Barnsley
          \inst{1}\fnmsep\thanks{rmb@astro.livjm.ac.uk}
          \and
          I.A.Steele\inst{1}
          }

   \institute{Astrophysics Research Institute, Liverpool John Moores University, Twelve Quays House, Birkenhead, CH41 1LD, U.K.}

   \date{Received September 09, 2012; accepted October 10, 2012}

 
  \abstract
   {}
   {We attempt to determine if a dependency on spectral subtype or $v$sin$i$ exists for stars undergoing phase-changes between B and Be states, as well as 
   for those stars exhibiting variability in H$\alpha$ emission.}
   {We analyse the changes in H$\alpha$ line strength for a sample of 55 Be stars of varying spectral types and luminosity classes using five epochs of observations taken 
   over a ten year period between 1998 and 2010.}
   {We find i) that the typical timescale between which full phase transitions occur is most likely of the order of centuries, although no dependency on spectral 
    subtype or $v$sin$i$ could be determined due to the low frequency of phase-changing events observed in our sample, ii) that stars with earlier spectral types and 
    larger values of $v$sin$i$ show a greater degree of variability in H$\alpha$ emission over the timescales probed in this study, and iii) a trend of increasing 
    variability between the shortest and longest baselines for stars of later spectral types and with smaller values of $v$sin$i$.}
   {}


   \maketitle
%

\section{Introduction}

The emission lines displayed by Be stars are known to be transient \citep{1990Ap&SS.163....7K}, with emission varying over timescales ranging from less than a day 
\citep{2002PASP..114..551P} to years \citep{2002PASP..114.1226M}. This emission variability is reflected in the first working definition given by 
\citet{1987pbes.coll....3C} who defined a Be star as ``a non-supergiant B star whose spectrum has, or had at some time, one or more Balmer lines in emission''. 

More recently, this definition has had to be refined to exclude emission-line stars whose disk formation mechanisms were accretion, e.g. Herbig AeBe stars. 
Consequently, a new subgroup of these emission-line stars, termed ``classical Be stars'', was created. Classical Be stars are characterised by their rapid rotation 
and decretion disks \citep[see][]{2003PASP..115.1153P}. \citet{2011IAUS..272..242M} proposed a refinement of 
Collins' definition to ``$\ldots$ a star with innate or acquired very fast rotation which combined to other mechanisms $\ldots$ leads to episodic matter ejections 
creating a circumstellar decretion disk or envelope'' to not only reflect the larger range of spectral types over which the phenomena have been observed to occur 
\citep{1974ApJ...193..113C, 2007IBVS.5773....1R} albeit with lesser frequency \citep{2004AN....325..749N}, but also to constrain the method whereby the disk 
is formed. It is these classical Be stars that are the subject of the following study.

In 1999, a multiwavelength survey of a representative sample of 58 Be stars was undertaken 
\citep{1999A&AS..137..147S, 2000A&AS..141...65C, 2001A&A...371..643S, 2001A&A...369...99H}. The sample is ``representative'' in that it attempted to include objects of 
each spectral type and luminosity class and so doesn't serve to reflect the spectral type or luminosity class distribution of Be stars. 

One of the conclusions drawn from this survey was that at the time of observation, non-emission line stars in the sample (i.e. by definition, ones that must have 
shown emission in the past) had lower rotational velocities and earlier spectral types, an implication being that they may be more prone to B/Be phase changes
\citep{2001A&A...371..643S}. Using optical spectrographic follow up taken ten years after the sample was first observed, the validity of these statements will be 
investigated further in this paper.


\section{Observations and data reduction}

The targets observed were taken from \citet[][hereafter S99]{1999A&AS..137..147S}. The catalogue names of the targets and some of their fundamental properties 
(viz. spectral class, luminosity class, V band mag., $v$sin$i$) are shown in Table \ref{tab:results}. 

\begin{table*}[p]
  \begin{center}
    \caption{The sample targets with their corresponding stellar parameters taken from S99. ``sh'' indicates that shell lines are observed in the spectra. 
    The measured equivalent widths of the sample H$\alpha$ lines are shown for each epoch. Missing observations are marked with an ellipsis.}
    \begin{tabular}{l l l l l l l l l l}
      \hline
      & & & & & & & & & \\ [-1ex]
      Object Name & Spec. Type & Lum. Class & $V$ Band & $v$sin$i$ & \multicolumn{5}{l}{EW} \\ [0.2ex]
      & & & & & INT & F1 & F2 & F3 & F4 \\ [0.2ex]
      & & & & (km/s) & & & & & \\ [0.2ex]
      & & & & & & & & & \\ [-1.5ex]
      \hline
      & & & & & & & & & \\ [-1ex]
      CD -28 14778  	&	 B2 	&	 III 	&	 8.95 	&	 153 	&	-32.9	&	\dots	&	\dots	&	-31.0	&	\dots	\\
      CD -27 11872 	&	 B0.5 	&	 V-III 	&	 8.69 	&	 224 	&	\dots	&	-34.9	&	\dots	&	-37.5	&	\dots	\\
      CD -27 16010 	&	 B8 	&	 IV 	&	 4.20 	&	 187 	&	0.0	&	-1.6	&	-1.0	&	-2.0	&	-1.1	\\
      CD -25 12642 	&	 B0.7 	&	 III 	&	 9.00 	&	 77 	&	2.7	&	2.5	&	\dots	&	2.6	&	\dots	\\
      CD -22 13183 	&	 B7 	&	 V 	&	 7.90 	&	 174 	&	-8.3	&	-14.2	&	\dots	&	-14.5	&	\dots	\\
      BD -20 05381 	&	 B5 	&	 V 	&	 7.80 	&	 202 	&	-12.1	&	-13.3	&	\dots	&	-14.6	&	\dots	\\
      BD -19 05036 	&	 B4 	&	 III 	&	 7.92 	&	 121 	&	4.5	&	4.6	&	\dots	&	\dots	&	\dots	\\
      BD -02 05328 	&	 B7 	&	 V 	&	 6.22 	&	 151 	&	-6.3	&	-5.2	&	\dots	&	-5.8	&	-6.3	\\
      BD -01 03834 	&	 B2 	&	 IV 	&	 8.14 	&	 168 	&	-35.9	&	-50.1	&	\dots	&	-45.2	&	\dots	\\
      BD -00 03543 	&	 B7 	&	 V 	&	 6.88 	&	 271 	&	-1.8	&	-5.0	&	\dots	&	-6.4	&	\dots	\\
      BD +02 03815 	&	 B7.5 	&	 sh 	&	 6.92 	&	 224 	&	-5.5	&	-6.2	&	\dots	&	-7.4	&	\dots	\\
      BD +05 03704 	&	 B2.5 	&	 V 	&	 6.13 	&	 221 	&	0.3	&	4.0	&	\dots	&	3.8	&	\dots	\\
      BD +17 04087 	&	 B6 	&	 III-V 	&	 10.6 	&	 156 	&	3.9	&	4.2	&	\dots	&	3.4	&	\dots	\\
      BD +19 00578 	&	 B8 	&	 V 	&	 5.69 	&	 240 	&	4.0	&	5.4	&	5.7	&	4.9	&	4.6	\\
      BD +20 04449 	&	 B0 	&	 III 	&	 8.3 	&	 81 	&	2.4	&	4.0	&	3.4	&	3.2	&	3.0	\\
      BD +21 04695 	&	 B6 	&	 III-V 	&	 5.78 	&	 146 	&	-0.7	&	\dots	&	-14.2	&	-15.3	&	\dots	\\
      BD +23 01148 	&	 B2 	&	 III 	&	 7.36 	&	 101 	&	0.6	&	\dots	&	0.7	&	\dots	&	1.1	\\
      BD +25 04083 	&	 B0.7 	&	 III 	&	 8.94 	&	 79 	&	2.5	&	2.5	&	3.3	&	2.5	&	2.4	\\
      BD +27 00797 	&	 B0.5 	&	 V 	&	 9.86 	&	 148 	&	-18.2	&	\dots	&	\dots	&	-29.7	&	-22.7	\\
      BD +27 00850 	&	 B1.5 	&	 IV 	&	 9.38 	&	 112 	&	4.0	&	\dots	&	4.2	&	3.1	&	2.7	\\
      BD +29 03842 	&	 B1 	&	 II 	&	 10.11 	&	 91 	&	1.6	&	2.4	&	\dots	&	1.7	&	1.7	\\
      BD +29 04453 	&	 B1.5 	&	 V 	&	 8.10 	&	 317 	&	-54.3	&	-61.7	&	-60.1	&	-58.8	&	-63.8	\\
      BD +30 03227 	&	 B4 	&	 V 	&	 6.58 	&	 218 	&	1.8	&	5.6	&	5.6	&	6.0	&	\dots	\\
      BD +31 04018 	&	 B1.5 	&	 V 	&	 7.16 	&	 211 	&	-18.6	&	-21.1	&	\dots	&	-19.6	&	-20.9	\\
      BD +36 03946 	&	 B1 	&	 V 	&	 9.20 	&	 186 	&	-23.2	&	-31.7	&	\dots	&	-35.7	&	-35.6	\\
      BD +37 00675 	&	 B7 	&	 V 	&	 6.16 	&	 207 	&	-12.7	&	-14.0	&	-12.2	&	-11.0	&	-11.2	\\
      BD +37 03856 	&	 B0.5 	&	 V 	&	 10.20 	&	 104 	&	3.6	&	3.4	&	\dots	&	3.3	&	2.7	\\
      BD +40 01213 	&	 B2.5 	&	 IV 	&	 7.34 	&	 128 	&	-12.2	&	\dots	&	-14.6	&	-17.9	&	-17.3	\\
      BD +42 01376 	&	 B2 	&	 V 	&	 7.28 	&	 196 	&	-7.8	&	-15.2	&	-15.9	&	-12.8	&	-13.9	\\
      BD +42 04538 	&	 B2.5 	&	 V 	&	 8.02 	&	 282 	&	-37.5	&	-37.8	&	-38.8	&	-41.6	&	-41.4	\\
      BD +43 01048 	&	 B6 	&	 III-sh &	 9.53 	&	 220 	&	-6.0	&	\dots	&	-8.8	&	-10.3	&	-9.5	\\
      BD +45 00933 	&	 B1.5 	&	 V 	&	 8.06 	&	 148 	&	3.9	&	3.5	&	3.7	&	3.7	&	\dots	\\
      BD +45 03879 	&	 B1.5 	&	 V 	&	 8.48 	&	 193 	&	-31.1	&	-23.4	&	-20.6	&	-22.7	&	-25.0	\\
      BD +46 00275 	&	 B5 	&	 III 	&	 4.25 	&	 113 	&	1.0	&	0.9	&	0.8	&	1.5	&	2.4	\\
      BD +47 00183 	&	 B2.5 	&	 V 	&	 4.50 	&	 173 	&	-29.7	&	-36.1	&	\dots	&	-38.0	&	-37.2	\\
      BD +47 00857 	&	 B4 	&	 V-IV 	&	 4.23 	&	 212 	&	-36.3	&	-38.7	&	-38.6	&	-43.0	&	-41.1	\\
      BD +47 00939 	&	 B2.5 	&	 V 	&	 4.04 	&	 163 	&	-22.4	&	-25.6	&	-24.5	&	-25.4	&	-26.1	\\
      BD +47 03985 	&	 B1.5 	&	 sh 	&	 5.42 	&	 284 	&	-17.8	&	\dots	&	-23.2	&	-25.3	&	\dots	\\
      BD +49 00614 	&	 B5 	&	 III 	&	 7.57 	&	 90 	&	-4.9	&	\dots	&	-6.4	&	-7.5	&	-7.0	\\
      BD +50 00825 	&	 B7 	&	 V 	&	 6.15 	&	 187 	&	-1.5	&	-2.1	&	-1.7	&	-2.6	&	-2.1	\\
      BD +50 03430 	&	 B8 	&	 V 	&	 7.02 	&	 230 	&	-6.2	&	-9.8	&	-9.7	&	-10.6	&	-10.9	\\
      BD +51 03091 	&	 B7 	&	 III 	&	 6.19 	&	 106 	&	2.2	&	1.0	&	\dots	&	0.7	&	1.1	\\
      BD +53 02599 	&	 B8 	&	 V 	&	 8.08 	&	 191 	&	-0.7	&	1.2	&	1.8	&	1.8	&	1.8	\\
      BD +55 00552 	&	 B4 	&	 V 	&	 7.90 	&	 292 	&	-1.3	&	-9.4	&	-10.4	&	-10.8	&	-10.6	\\
      BD +55 00605 	&	 B1 	&	 V 	&	 9.34 	&	 126 	&	-5.9	&	-7.5	&	-5.2	&	2.2	&	-0.7	\\
      BD +55 02411 	&	 B8.5 	&	 V 	&	 5.89 	&	 159 	&	2.9	&	2.1	&	2.7	&	2.9	&	2.7	\\
      BD +56 00473 	&	 B1 	&	 V-III 	&	 9.08 	&	 238 	&	-37.4	&	-18.2	&	-20.1	&	\dots	&	-17.3	\\
      BD +56 00478 	&	 B1.5 	&	 V 	&	 8.51 	&	 157 	&	-9.0	&	-12.5	&	-13.5	&	-16.8	&	\dots	\\
      BD +56 00484 	&	 B1 	&	 V 	&	 9.62 	&	 173 	&	-47.8	&	-46.6	&	-47.2	&	-44.6	&	-36.7	\\
      BD +56 00493 	&	 B1 	&	 V-IV 	&	 9.62 	&	 270 	&	2.0	&	-0.3	&	1.0	&	-2.4	&	-1.9	\\
      BD +56 00511 	&	 B1 	&	 III 	&	 9.11 	&	 99 	&	-5.1	&	-4.2	&	-5.4	&	-6.4	&	-6.1	\\
      BD +56 00573 	&	 B1.5 	&	 V 	&	 9.66 	&	 250 	&	-67.5	&	-38.5	&	-73.3	&	-77.1	&	-74.6	\\
      BD +57 00681 	&	 B0.5 	&	 V 	&	 8.66 	&	 147 	&	0.2	&	1.4	&	1.6	&	0.6	&	\dots	\\
      BD +58 00554 	&	 B7 	&	 V 	&	 9.16 	&	 229 	&	-9.5	&	-7.9	&	-7.6	&	-8.9	&	\dots	\\
      BD +58 02320 	&	 B2 	&	 V 	&	 9.51 	&	 243 	&	-4.1	&	-3.9	&	-5.2	&	-5.9	&	-6.0	\\
      & & & & & & & & & \\ [-1ex]
      \hline
      \\
    \end{tabular}
    \label{tab:results}
  \end{center}
\end{table*}

\afterpage{\clearpage}

In S99, the stars were reclassified under the MK scheme \citep{1973ARA&A..11...29M} using spectra taken by the Intermediate Dispersion Spectrograph (IDS) on the Isaac 
Newton Telescope (INT), rather than relying on values taken from the literature. Rotational velocities were also determined in S99 by fitting Gaussian profiles to 4 
HeI lines at 4026\AA, 4143\AA, 4387\AA \, and 4471\AA \, and applying the full width half maximum - $v$sin$i$ correlation of \citet{1975ApJS...29..137S}. It should be 
noted that this method of inferring a $v$sin$i$ has not taken into account the effect of gravity darkening, which has been shown to introduce a redundancy between 
line profile width and $v$sin$i$ at the largest rotation speeds \citep{2004MNRAS.350..189T, 2005A&A...440..305F}, leading to underestimates of the true rotation 
speed for the fastest rotators. Residual emission within the HeI lines may have also introduced a bias towards lower $v$sin$i$ for earlier subtypes. 

Five epochs of observations were used in the following analysis (hereafter referred to as the INT, F1, F2, F3 and F4 datasets) and were obtained using both the IDS 
and FRODOSpec \citep{2004AN....325..215M} on the Liverpool Telescope \citep{2004SPIE.5489..679S}.

The IDS observations used in this analysis were made on the night of 1998 August 3 using the R1200Y grating with a slit width of 1.15 arcsec, corresponding to
a dispersion of $\sim$0.5 \AA/pixel on the EEV12 CCD. A central wavelength of 6560 \AA \,was chosen, giving a wavelength range between 5800 -- 7100 \AA.

Reduction of INT data was performed using Figaro according to the standard prescription for long-slit spectra. Tungsten lamp flats were used to 
map the slit response and sky flats the pixel-to-pixel flat field variations. Target spectra were extracted using simple extraction and a sky region of the same size 
was also extracted and subtracted from each target spectrum. To wavelength calibrate the target frames, arc exposures were taken at the start, middle and end of the 
night using CuAr and CuNe lamps. After extracting their spectra, the calibrations were determined and copied onto the target spectra. 

All FRODOSpec observations were taken over a period between September 2009 and November 2010. Observations for the F1 dataset were made between 08/09/2009 and 26/09/2009, F2 
between 03/11/2009 and 26/12/2009, F3 between 18/07/2010 and 29/08/2010 and F4 between 10/10/2010 and 23/11/2010. 

The first two epochs of data, F1 and F2, were taken using the red 
diffraction gratings. The last two, F3 and F4, were taken using the higher resolution red VPH gratings. FRODOSpec provides wavelength coverage from
3900--5700\AA \, (blue arm) and 5800--9400\AA \, (red arm) for the lower resolution configuration (R = 2600 and 2200 respectively) and 3900--5100\AA \, (blue arm) and 
5900--8000\AA \, (red arm) for the higher resolution (R = 5500 and 5300 respectively). FRODOSpec reduction was performed using the pipeline discussed in 
\citet{2012AN....333..101B}.

%

\section{Method}

Emission in the Balmer series is commonly used to provide an insight into the circumstellar environment surrounding Be stars \citep{2006ApJ...651L..53G} and results 
from recombination of photospheric radiation within the disk. In this paper, only the line strength of H$\alpha$ is measured. H$\alpha$ is 
observed in Be stars with a variety of profile shapes \citep{2000A&AS..147..229B}, ranging from single and double peaked emission to absorption, where no 
identifiable emission component can be detected (i.e. the star is not currently in an emission phase). 

\subsection{Quantifying the extent of disk loss/formation}

Due to in-filling of photospheric absorption lines, visual inspection alone is not reliable at being able to discern between a star that shows weak emission and a 
star showing no emission at all, especially when the observations vary in resolution. It was therefore necessary to use a more quantitative method, capable of 
quantifying the extent of disk emission in a more robust manner. For this reason, the equivalent width (EW) of the line was used to both assess phase change and quantify 
the variability of the emission.

In order to measure the EWs, the spectra were first normalised by fitting a polynomial to the shape of the underlying continuum. The EW was then calculated 
over a suitable wavelength range encompassing the line. The EW measurement was performed in IDL using the \texttt{line\_eqwidth} routine from the ltools 
package\footnote{http://fuse.pha.jhu.edu/analysis/fuse\_idl\_tools.html}. 

The dominant source of error in measuring the EW is the determination of the continuum level and subsequent normalisation \citep{1538-3881-141-5-150}, with any 
scaling error in determining the level of the continuum leading directly to a multiplicative error in the EW. To obtain an empirical estimate of this uncertainty, 
sets of repeated observations of B stars with different brightnesses spanning the sample range  (V = 5, 8.5 and 10)
were taken. These observations were taken sequentially on the same night with both FRODOSpec's resolution configurations, using corresponding exposure times similar to those 
used to observe the sample.

The standard deviation of the measured H$\alpha$ absorption EWs was calculated for each set of target spectra. The V band mags. and the measurement errors for each 
target are shown in Table \ref{tab:EWs_standards_repeated}.

These errors were associated to all measurements of EW, matching them to the nearest target brightness. In the absence of equivalent INT observations, IDS spectra were 
assigned the same error figures as those used for FRODOSpec. In analysis requiring the photospheric contribution to be subtracted off, the errors in measuring the EW 
and the error in assigning a photospheric EW were added in quadrature.

\begin{table}[ht]
  \begin{center}
    \caption{V band mags. and values of EW used as the continuum normalisation error.}
    \begin{tabular}{l l l}
      \hline
      & & \\ [-1ex]
      Object Name & $V$ & EW(H$\alpha$) $\sigma$ \\ [0.2ex]
      & & \\ [-1.5ex]
      \hline
      & & \\ [-1ex]
      HR 5834	& 5.00 & 0.30 \\
      HD 75750	& 8.48 & 0.1 \\
      HD 100340	& 10.12 & 0.1 \\
      & & \\ [-1ex]
      \hline
      \\
    \end{tabular}
  \label{tab:EWs_standards_repeated}
  \end{center}
\end{table}

Before EW measurements of the H$\alpha$ line could be used to assess if an object exhibited complete disk loss/formation, it was required that the 
photospheric contribution to the EW from the central star, $W_{phot}$, be subtracted off the measurement of the total EW of the line, $W_{tot}$, i.e.
\begin{equation}
  W_{disk} = W_{tot} - W_{phot}
\end{equation}
where $W_{disk}$ is the equivalent width of the emission line arising from the disk only. 

$W_{phot}$ can be established by either using models of stellar atmospheres \citep[e.g.][]{1979ApJS...40....1K} or by measuring the EW for a set of B stars (with
no history of emission) over a range of spectral types and luminosity classes and assuming a linear fit to EW and spectral type for each luminosity class. In order
to retain model independency in the following analysis, the latter method was prefered and a set of B stars, shown in Table \ref{tab:standards}, were 
observed.

\begin{table}[ht]
  \begin{center}
    \caption{The observed standards with their corresponding stellar parameters and measured H$\alpha$ equivalent widths.}
    \begin{tabular}{l l l l}
      \hline
      & & & \\ [-1ex]
      Object Name & Spec. Type & Lum. Class & EW \\ [0.2ex]
      & & & \\ [-1.5ex]
      \hline
      & & & \\ [-1ex]
      HD 886	&	B2	&	IV	&	3.8	\\
      HD 20365	&	B3	&	IV	&	4.9	\\
      HD 22951	&	B0.5	&	V	&	4.2	\\
      HD 23180	&	B1	&	III	&	3.5	\\
      HD 23288	&	B7	&	IV	&	6.1	\\
      HD 23338	&	B6	&	V	&	5.7	\\
      HD 23850	&	B8	&	III	&	5.8	\\
      HD 29248	&	B2	&	III	&	3.4	\\
      HD 147394	&	B5	&	IV	&	5.3	\\
      HD 160762	&	B3	&	IV	&	4.1	\\
      HD 176437	&	B9	&	III	&	6.9	\\
      HD 184930	&	B5	&	III	&	4.9	\\
      HD 186882	&	B9.5	&	III	&	7.6	\\
      HD 196867	&	B9	&	V	&	7.6	\\
      HD 207330	&	B2.5	&	III	&	3.9	\\
      HD 214923	&	B8	&	V	&	7.3	\\
      HD 214993	&	B1.5	&	III	&	4.0	\\
      HD 218376	&	B0.5	&	III	&	3.0	\\
      HD 219688	&	B5	&	V	&	4.5	\\
      HD 222661	&	B9.5	&	V	&	9.6	\\
      & & & \\ [-1ex]
      \hline
      \\
    \end{tabular}
  \label{tab:standards}
  \end{center}
\end{table}

\begin{figure}[ht]
  \begin{center}
  \includegraphics[scale=0.45]{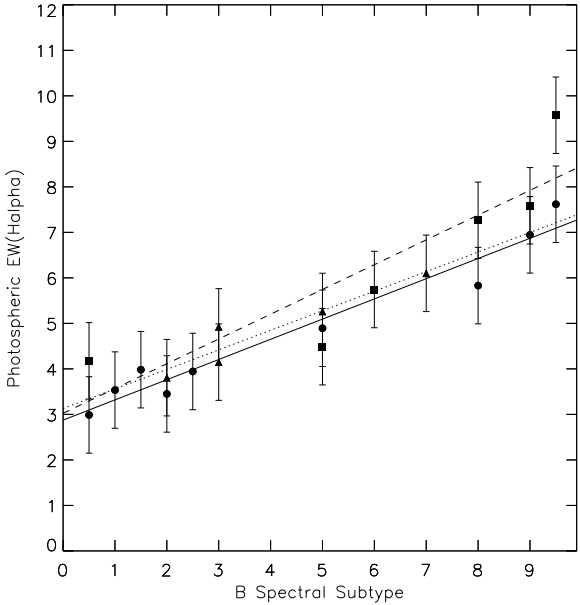}
  \caption{Photospheric EW as a function of spectral subtype for luminosity class III (circle), IV (triangle) and 
  V (square) with linear best-fit lines overplotted (solid, dotted and dashed for classes III, IV and V respectively).}
  \label{fig:standards_EW_spectype}
  \end{center}
\end{figure}

A plot of the EW against spectral subtype for each luminosity class is shown in Fig. \ref{fig:standards_EW_spectype}. The corresponding Pearson 
correlation coefficients for luminosity classes III, IV and V are -0.97, -0.94 and -0.88 respectively, indicating that the data are reasonably well described 
by a linear fit. 

As an aside, the measured EWs were compared with those determined from using Kurucz models. The model was found to underestimate for
earlier spectral types and overestimate for later, possibly due to the effects of line blending at the 20\AA \, resolution of the grid around the H$\alpha$ line.

%

\section{Results and Discussion}

The measured equivalent widths of the sample H$\alpha$ lines are shown for each epoch in Table \ref{tab:results}.

\subsection{Complete disk loss/formation}\label{ss:complete}
 
\begin{figure}[!t]
  \begin{center}
  \includegraphics[scale=0.45]{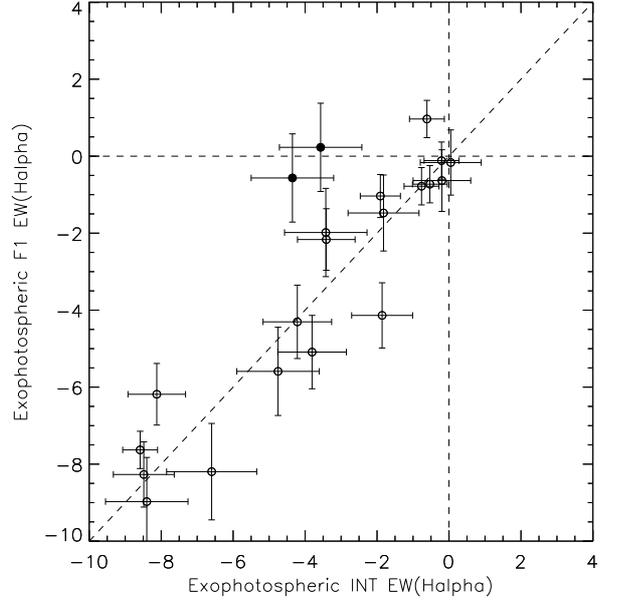}
  \caption{Determined exophotospheric EWs plotted for the INT-F1 baseline. The scale has been selected to show all targets that exhibited either disk loss or 
  formation. Targets that have been identified as losing their disk are illustrated with filled circles. The stability line (y=x) and disk loss/formation (y=0, x=0) axes are 
  shown.} \label{fig:HA_BYEW_COMPLETE_INT&F1}
  \end{center}
\end{figure}

To determine complete disk loss/formation using the calculated exophotospheric EWs, values for consecutive pairs of epochs (with corresponding baselines of
INT -- F1 ($\sim$10 years), F1 -- F2 ($\sim$2 months), F2 -- F3 ($\sim$9 months) and F3 -- F4 ($\sim$3 months)) were plotted against each other; see 
Fig. \ref{fig:HA_BYEW_COMPLETE_INT&F1} for an example plot. Candidates for 
disk loss were selected by identification of those targets that lay off the stability line (where no EW change could reliably be inferred), but whose EW and 
associated error lay within the disk loss (y=0) axis. Likewise, candidates for disk formation were selected using similar criteria, except only those that lay 
within the disk formation (x=0) axis were selected. The resulting phase-changers are shown with filled symbols in Fig. \ref{fig:HA_BYEW_COMPLETE_INT&F1}. As 
phase-changing was seen to occur only over the timescale of 10 years, only this baseline is shown.

One of the aims of this analysis was to assess if there was any correlation between spectral type and frequency of B/Be phase change. 
Because of the low number of targets that were identified undergoing this change, no attempt has been made to determine if a dependency on this 
fundamental parameter exists. Indeed, complete phase change could only be confirmed at the 3$\sigma$ level for two targets in the sample, BD +05 03704 and 
BD +30 03227, with the longest baseline of ten years (INT -- F1).

If these targets are believed to have truly changed state between B and Be phases, it is possible to estimate the approximate time over which complete 
phase change is expected to occur. As little observational evidence is available regarding both the distribution of phase-changing timescales and the fraction of 
Be stars which undergo complete phase change (most likely due to observational bias from incomplete coverage of the time domain), the following assumptions have been 
made:

\begin{itemize}
 \item The time between complete phase changes is randomly distributed (neither shorter nor longer times are preferred), and 
 \item All Be stars are subject to these episodic events of complete disk loss/formation.
\end{itemize}

With these in mind, a Monte Carlo simulation was set up to predict the number of observed phase-changers over a ten year period for different timescales 
(whilst constraining the minimum time to 10 years - the only baseline over which phase change was observed in this sample). The simulation was run iteratively using a 
55 star dataset, recording how many were observed to change phase an odd number of times within a defined ten year window (an even number would imply that no phase 
change would be observed). In a sample of 55 objects, the maximum time for phase-changing to yield 2 targets observed in a ten year baseline was found to be most 
likely between 500 and 1000 years.

However, this maximum time comes with a possible caveat. Since Be phenomena are time variable, selection of candidate Be stars
for the sample was based on heterogenous data taken from a multitide of different sources and observers that has been compiled over decades to produce a single
catalogue \citep{1982IAUS...98..261J}. If the original classification of the targets used in the sample is untrustworthy, the sample may be 
contaminated by B stars that have never truly shown reliable exophotospheric emission. This is certainly a real possibility for the fainter targets recorded 
in the catalogue, and to some extent may even be true for some of the brighter targets. 

If fewer true Be targets were present in the sample, the observed fraction of phase-changers to total targets increases, increasing the frequency  
of phase-changes over the timescales surveyed by this study and effectively decreasing the maximum time over which the targets are expected to change phase.
To determine the lower limit for the maximum time, the sample was filtered by-eye to remove targets that were always in absorption. As noted previously, due to 
in-filling of photospheric absorption lines this is not necessarily indicative that the star is not in emission, but serves as an upper limit to the number of stars 
that are non-variable, and thus could have been misclassified. The previous Monte Carlo simulation was repeated using the smaller dataset, and in this smaller sample 
of 44 objects, the maximum time for phase-changing to yield 2 targets observed in a ten year baseline was found to be most likely between 250 and 500 years. Finally, 
we note that this value will drop further if all Be stars are not subject to these episodic events of complete disk/loss formation.

\subsection{Partial disk variability}

In the following analysis, partial disk variability is conservatively defined as a 5$\sigma$ change in EW measurement, where the previous constraints placed to force 
the targets to lie upon either the axis of disk loss or formation are removed. As the underlying photospheric contribution from the central star will not change 
between epochs, the total EW is used in the analysis of partial disk variability. 

\subsubsection{As a function of spectral subtype}\label{sss:as_a_function_of_spectral_subtype}

\begin{figure}[ht]
  \begin{center}
  \includegraphics[scale=0.45]{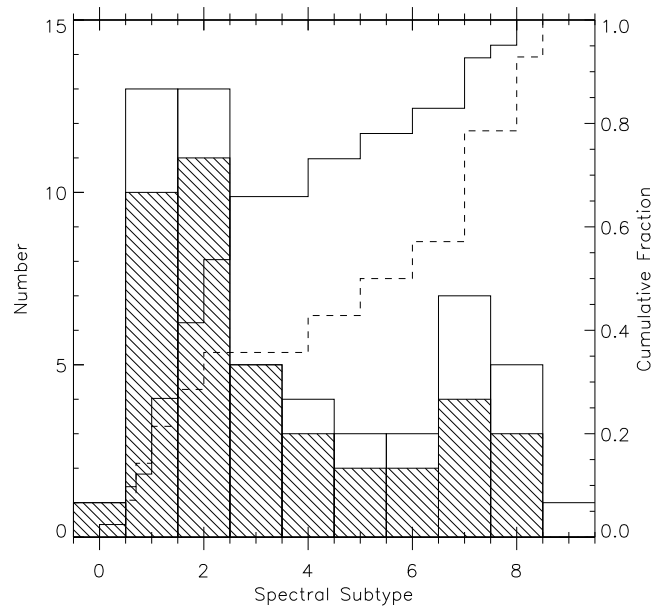}
  \caption{Targets showing partial variability over any of the baselines. The hollow histogram represents the distribution of targets as a function
  of spectral subtype. The hatched histogram shows the distribution of those targets displaying partial variability. The cumulative fractions
  have been overplotted for those targets displaying partial variability (solid line) and those not (dashed line).}
  \label{fig:HA_BYEW_ANY_BASELINE_SPECTYPE}
  \end{center}
\end{figure}

An initial approach to the analysis of partial variability is to assess its presence over all of the possible baselines. Specifically, 
if an object is seen to vary in at least one of the baselines, it should be classed as variable. A plot of the variability over any baseline as a function of spectral 
subtype is shown in Fig. \ref{fig:HA_BYEW_ANY_BASELINE_SPECTYPE}.

Analysis regarding variability as a function of spectral subtype is presented by \citet{1998A&A...335..565H} who found that in a homogenously obtained Hipparcos 
sample of 273 Be stars, $\ge$ 98\% of early Be stars (B0 -- B3e) showed some degree of variability, compared with only 45\% of late Be stars (B7 -- B9e) over the 
four years during which the survey was conducted. Using the same definition of early and late types, similar figures are found in this sample, where 84\% of the 
early Be stars in the sample were variable with only 46\% of the later type showing some variation at the 5$\sigma$ level over all baselines. 
\citet{1538-3881-141-5-150} find the same trend in their sample, but find 45\% of their early and 29\% of late types to be variable. These smaller fractions could be 
the result of the statistic they use to quantify variability, which they state as being ``sensitive to significant changes in disk density''.

There are several possible explanations for the observation of increased variability in earlier spectral types. It could be the case that a false-positive 
misclassification has introduced a bias, 
but such a bias would only alter the trend if later type non-variable B stars were preferentially misclassified as being variable. Filtering the sample by-eye to remove
targets that were always in absorption (as in \S \ref{ss:complete}), it is found that the trend remains with 100\% of the earlier type and 53\% of the later 
type varying. Non-variability was found for 7 out of the remaining 45 stars, with all 7 stars of spectral type B5 or later. 

Discounting this, the most obvious explanation would be that later spectral types are genuinely less variable than earlier. 
However, this may not necessarily be the complete picture, in so much that the timescales over which later types vary may not have been adequately probed by the 
timescales in this study.

To assess if there is any trend in the degree of variability across different timescales, pairs of epochs were grouped together to construct sets of similar baseline 
timescales. For the purposes of the following analysis, the baselines have been separated into $<$ 1 year (F1-F2, F3-F4), about one year (F1-F3, F2-F3, F1-F4, F2-F4) 
and ten years (INT-F1, INT-F2, INT-F3, INT-F4). The previous analysis was then redone using the each of the sets; the result is shown in Fig. 
\ref{fig:HA_BYEW_BASELINE_SETS_SPECTYPE}. The following figures are presented for both the full dataset and the previous emission-only filtered dataset 
(given in brackets after the full dataset).

\begin{figure}[ht]
  \begin{center}
  \includegraphics[scale=0.45]{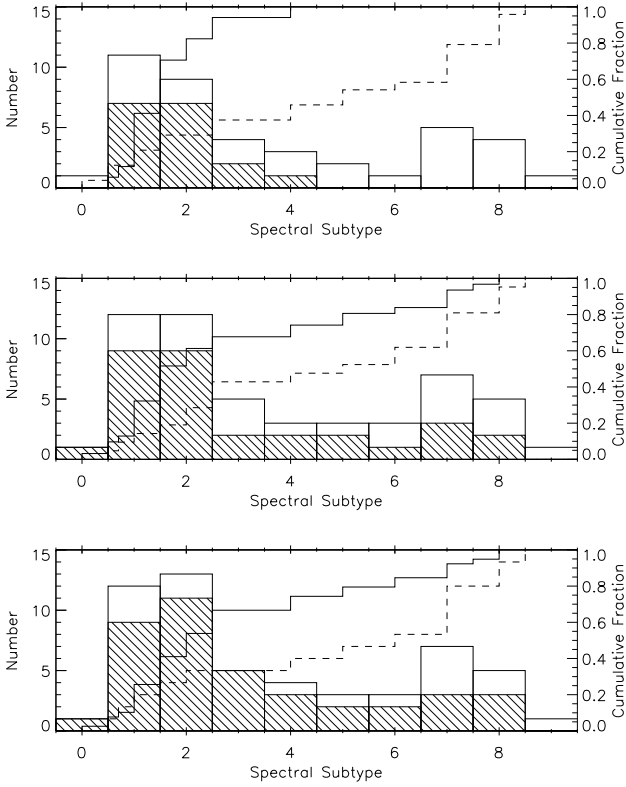}
  \caption{Targets showing partial variability over $<$1 yr (top), $\sim$1 yr (middle) and $\sim$10 yrs (bottom). The hollow histograms 
  represent the distribution of targets as a function of spectral subtype. The hatched histograms show the distribution of those targets displaying partial 
  variability. The cumulative fractions have been overplotted for those targets displaying partial variability 
  (solid line) and those not (dashed line).}
  \label{fig:HA_BYEW_BASELINE_SETS_SPECTYPE}
  \end{center}
\end{figure}

For partial variability over timescales less than a year, no later types could be confirmed at the 5$\sigma$ level to be partially varying in either dataset, whilst 
64\% (79\%) of the earlier type were. For timescales of around 10 years, the number of later types varying were found to be 46\% in both datasets, with the number of 
earlier types also rising to 84\% (100\%). This would seem to imply some dependency of partial variability as a function of spectral type on baseline, but there is 
little difference in the distributions between timescales of one year and timescales of 10 years. It is worth noting that as the distribution of Be stars peaks at 
around a spectral type of B2 \citep{2010A&A...509A..11M} and tails off sharply towards later spectral types, quantitative statistics derived using these later objects 
are less certain.

A two sample Kolmogorov-Smirnov (K-S) test was also used to determine if the observed distribution of targets showing partial variability differed from the underlying 
parent distribution of those which were non-variable as a function of spectral type. The calculated rejection probability for the same distribution 
hypothesis was found to be 75\% (99\%), the implication being that the stars displaying partial variability may have a different parent distribution, and thus 
dependency on spectral type, than those which don't.

\subsubsection{As a function of rotational velocity}

\begin{figure}[ht]
  \begin{center}
  \includegraphics[scale=0.45]{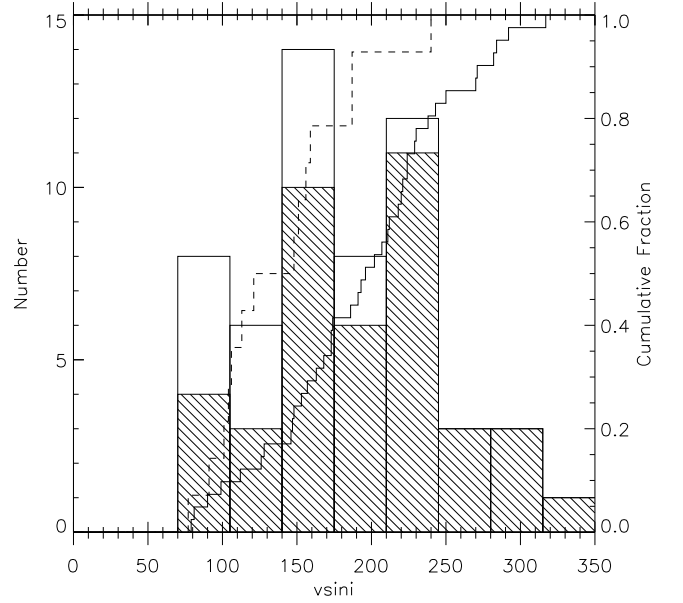}
  \caption{Targets showing partial variability over any of the baselines. The hollow histogram represents the distribution of targets as a function
  of $v$sin$i$. The hatched histogram shows the distribution of those targets displaying partial variability. The cumulative fractions
  have been overplotted for those targets displaying partial variability (solid line) and those not (dashed line).}
  \label{fig:HA_BYEW_ANY_BASELINE_VSINI}
  \end{center}
\end{figure}

A plot of the variability over any baseline as a function of $v$sin$i$ is shown in Fig. \ref{fig:HA_BYEW_ANY_BASELINE_VSINI}. From visual 
inspection of the plot, it would appear that stars with lower rotational velocity are less variabile over the timescales probed by this study.

The two sample K-S test was also used to determine if the observed distribution of targets showing partial variability differed from the underlying 
parent distribution of those which were non-variable as a function of $v$sin$i$. The calculated rejection probability for the same distribution 
hypothesis was found to be $>$99\% (81\%), the implication being that the stars displaying partial variability may have a different parent distribution, 
and thus dependency on $v$sin$i$, than those which don't. 

\begin{figure}[ht]
  \begin{center}
  \includegraphics[scale=0.45]{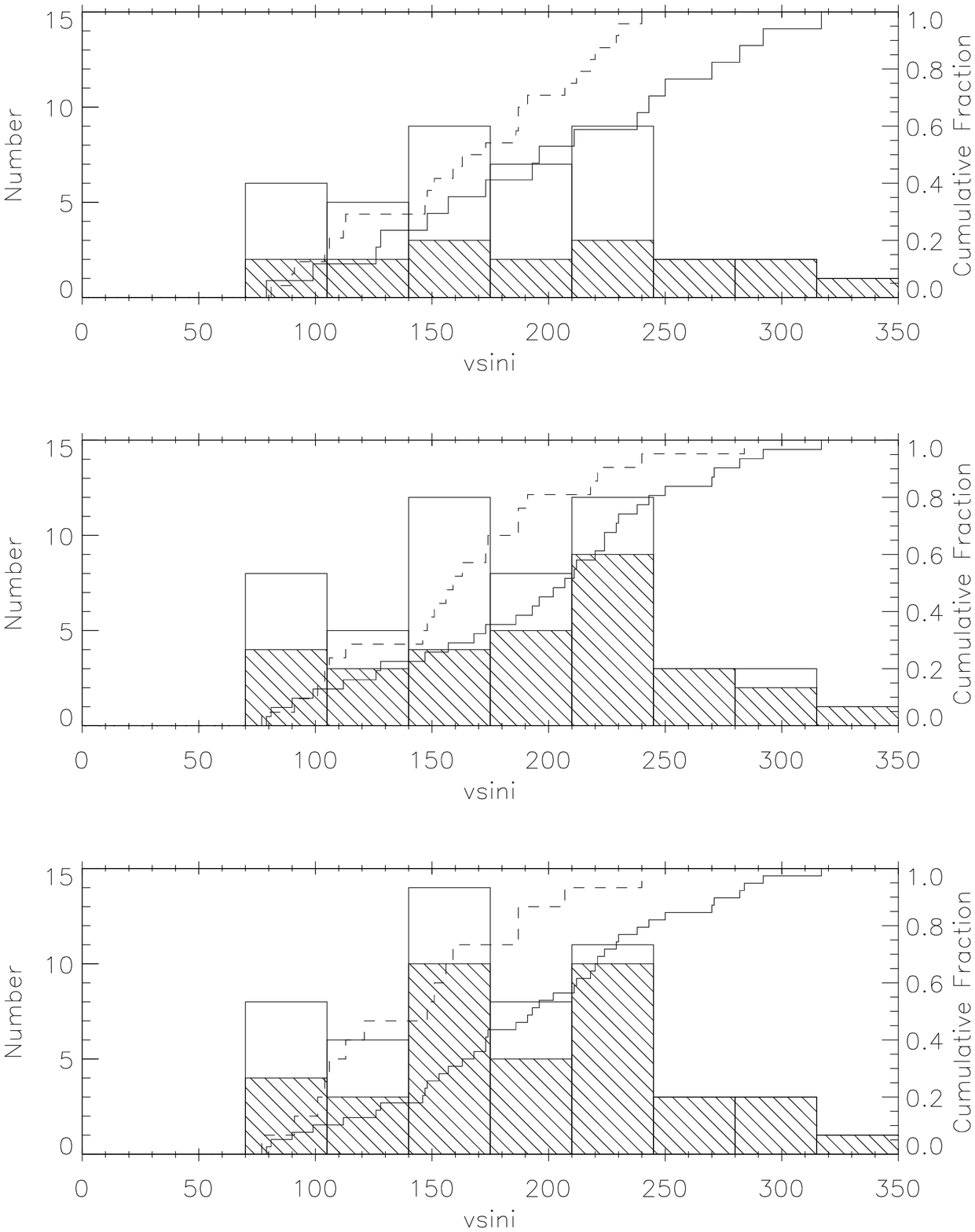}
  \caption{Targets showing partial variability over $<$1 yr (top), $\sim$1 yr (middle) and $\sim$10 yrs (bottom). The hollow histograms 
  represent the distribution of targets as a function of $v$sin$i$. The hatched histograms show the distribution of those targets displaying partial 
  variability. The cumulative fractions have been overplotted for those targets displaying partial variability 
  (solid line) and those not (dashed line).}
  \label{fig:HA_BYEW_BASELINE_SETS_VSINI}
  \end{center}
\end{figure}

For completeness, the trend of variability as a function of $v$sin$i$ is also presented using the same epoch pairings discussed in 
\S \ref{sss:as_a_function_of_spectral_subtype}; the result is shown in Fig. \ref{fig:HA_BYEW_BASELINE_SETS_VSINI}. Partial variability 
is seen to occur for nearly 100\% of those targets with the highest $v$sin$i$ over all baselines, with a trend of increasing variability for longer baselines seen 
for stars with lower $v$sin$i$. 

However, it is noted that analysis using $v$sin$i$ may be misleading. To compare stars using this quantity is to compare stars which may have different inclination 
angles (the angle between the rotation axis and the line of sight), spectral types and luminosity classes. A more useful parameter is the ratio of the observed 
rotational velocity to the critical, or break-up, velocity of the star (which in itself is a function of spectral type and luminosity class) given by 
$\omega = v_{rot}/v_{crit}$. However, a model-independent determination of $v_{crit}$ is not attainable, and as such we do not attempt to perform analysis using 
this quantity.



%

\section{Conclusions}

\begin{itemize}
 \item The observed frequency at which transitions between the B and Be states occur is too low (2/55) for the purposes of ascertaining any dependency on spectral 
 type or $v$sin$i$. Given this, we predict that the typical timescale between which these complete phase transitions occur is likely to be of the order of centuries.
 \item Using a measure of spectral type dependency akin to that used by previous authors, we find evidence to suggest that stars of an earlier subtype may
 be more variable. We also find a rise in variability for later types (B7-9) when observed over baselines greater than a year. However, the weight with which this 
  conclusion is made is limited by the small number of stars observed at later subtypes.
 \item Similar to the previous conclusion, we find that stars with larger values of $v$sin$i$ may be more variable. Again we note a rise in variability for stars with 
  lower values when observed over baselines greater than a year. 
 \item The distributions of variable and non-variable stars with respect to both spectral type and $v$sin$i$ are found to be different under a two sample 
  K-S test.
\end{itemize}

\begin{acknowledgements}
The Liverpool Telescope is operated on the island of La Palma by Liverpool John Moores University in the Spanish Observatorio del Roque de los Muchachos of the Instituto de 
Astrofisica de Canarias with financial support from the UK Science and Technology Facilities Council. RMB acknowledges STFC for a postgraduate studentship during 
which this work was undertaken. We thank the anonymous referee whose detailed review has helped to improve this paper. 
\end{acknowledgements}

\bibliographystyle{aa}
\bibliography{rmb_be1}

\end{document}